\documentclass[onecolumn, prl]{revtex4}

\usepackage{graphicx,amssymb, amsmath}
\usepackage{verbatim}
\usepackage{times}

\newcommand{\K}{{\vec{k}}}
\newcommand{\R}{{\vec{r}}}
\newcommand{\RR}{{\vec{R}}}
\newcommand{\En}{{\omega}}

\begin{document}

\title{Extracting Quasiparticle Lifetimes from STM experiments}
\author{Sumiran Pujari}
\affiliation{Department of Physics,
Cornell University, Ithaca, New York 14853-2501}
\email{sp384@cornell.edu}

\begin{abstract}
Based on Quasiparticle interference(QPI) around a point impurity, we demonstrate an analysis scheme that extracts
the lifetime of a quasiparticle by using the local density of states(LDOS) data around the 
impurity in a Scanning Tunneling Microscopy(STM) experiment. This data analysis scheme would augment 
the Fourier-Transform Scanning Tunneling Spectroscopic
methods which provides us with the quasiparticle dispersion. Thus, point impurities can be \emph{used} as probes to extract 
quasiparticle lifetimes from STM experiments and this would complement other experimental methods 
such as Angle Resolved Photo-emission
Spectrocopy(ARPES). We detail how the scheme would apply to metals and superconductors. 
\end{abstract}

\maketitle

Scanning Tunneling Microscopy(STM) has revolutionized condensed matter research by providing us with
unprecedented detail on the local real space electronic properties of the sample under investigation. 
But even more remarkably, it has been shown that momentum space properties of the sample can be extracted through
the application of Fourier Transform Scanning Tunneling Spectroscopy (FT-STS) \cite{Sprunger-Petersen}. 
Using FT-STS one figures
out the dispersion of the underlying well-defined quasiparticles or carriers. 

In this paper, we aim to extend the domain of momentum space properties that can be extracted using STM.
The central result is  the demonstration of a data analysis scheme that would give us the lifetimes
	of the charge carriers in a sample as a function of momentum(and energy)
from data collected in an STM experiment. Previously, ARPES is the tool that has been used successfully
to extract lifetimes of carriers in a sample by measuring the one-particle electron spectral
function directly in momentum space. Extracting lifetime information from STM - a real space probe - thus would add value 
by providing an independent method that complements and checks the ARPES method. Previous attempts at
reconciling lifetime broadening effects on STM data mainly consist of writing down viable fitting forms for the 
lifetime function that fit with the STM data, rather than extracting it out of the data directly
 like one does in an ARPES experiment by
quantifying the width of the peaks in ARPES spectra (See e.g. \cite{Valla}). In the context of metals/Fermi liquids,
Ref. \cite{Burgi} have fitted STM data on Silver and Copper with a model for thermal broadening of the
electrons \cite{Adawi}. Refs. \cite{YRZ} and \cite{Jacob} are prominent examples in the STM phenomenology of
high temperature superconductors.

We start by describing
the scheme in the simpler case of normal metals. We imagine the system to be
composed of Landau quasiparticles described by a propagator of the form
\begin{equation}
\tilde{G}_0(k;\omega) = \frac{1}{\omega - i \eta(k,\omega) - \epsilon(k)}
\label{eq:FLprop}
\end{equation}
Self-energy processes - e.g. due to electron-electron interaction as in a Fermi Liquid
or through scattering off a bosonic mode like phonons -
lead to a finite lifetime for the quasiparticle and this is formally taken care by the imaginary term in the
denominator of Eq. (\ref{eq:FLprop}), $\eta(k,\omega)$. Also, the real part of the self-energy shifts the 
chemical potential and we assume that the dispersion term $\epsilon(k)$ is this shifted 
dispersion \cite{shift}. Our 
aim is to extract $\eta(k,\omega)$ from STM data. We assume the knowledge of the dispersion
$\epsilon(k)$ either via FT-STS on the same data or through an ARPES experiment. 

Apart from the quasiparticles, let us imagine there to be a point impurity in the system, say at origin, which scatters the 
quasiparticles. In a real situation, we are imagining there to be a dilute amount of impurities in the sample 
so that multiple impurity scattering is not important. For a single point impurity,
The impurity problem is solved via the T-matrix approach \cite{Rickayzen}, and the real space impurity scattered
electron propagator is given by
\begin{equation}
G(r,r';\omega) = G_0(r,r';\omega) + G_0(r,r_{imp};\omega) \cdot T(\omega) \cdot G_0(r_{imp},r';\omega)
\label{eq:Tmatrix}
\end{equation}
where $G_0(r,r';\omega) = (2 \pi/L)^2 \sum_k \tilde{G}_0(k;\omega) e^{i k.(r-r')} \equiv G_0(R=r-r';\omega)$
is the free electron propagator and the impurity effect 
is captured by the so-called ``T-matrix" $T(\omega)$, which is given by $T(\omega) = U/(1 - U G_0(r_{imp},r_{imp}
;\omega))$ where $U$ is the impurity strength. It is in the second term of the above equation that we have
QPI which is utilized in FT-STS. 

We will quickly review the key notions underlying FT-STS, since our method also utilizes QPI. 
STM measures LDOS as a spatial map over the surface for a range of energies. The LDOS $n(r;\omega)$
is proportional to imaginary part of the real space propagator, i.e. 
\begin{equation}
	n(r;\omega) = -\frac{1}{\pi} Im [G(r,r;\omega)].
\label{eq:LDOS}
\end{equation}
FT-STS's main operating principle is that
the peaks in the Fourier transform of LDOS map at a particular energy
are at wave-vectors which connect pairs of points on the $\epsilon(k)$'s contour at that particular energy 
for which the joint density of states is maximum. This can be understood  
by looking at the Fourier transform of the interference term in Eq. (\ref{eq:Tmatrix}) (see Eq. (1)
of \cite{WangLee} and the following paragraph). If the quasiparticles have finite lifetimes, 
the resultant effect in FT-STS will
be a broadening of the FT-STS peaks(which are seen in experiments, e.g. \cite{SeamusQPI}). Moreover, the ``shapes"
of these FT-STS peaks contain information about the momentum dependence of the lifetime $\eta(k;\omega)$. 
It seems that extracting the $k$-dependence of $\eta(k;\omega)$ from the FT-STS method is a hard task because,
apart from other possible broadening factors like inhomogeneity (e.g. STM on cuprates), 
one has the difficulty of deconvolving the output of
FT-STS - the QPI term is a product in real space -
without the prior knowledge of $\eta(k;\omega)$.
Instead we will work in real space,
our main tactic being to extract $G_0(R;\omega)$ from QPI,
and STM data is most suited for this.  

We now list down the main steps of the analysis scheme and in what follows we give their essential technicalities
along with pictorial demonstrations. In the Appendix, we include further technical details and proofs required in 
those steps. 1) From LDOS/$n(r;\omega)$ map, we construct a $G(r,r;\omega)$ map. 2) Once we have the
$G(r,r;\omega)$ data, we ``invert" Eq. (\ref{eq:Tmatrix}) in order to extract $G_0(R;\omega)$.
To invert Eq. (\ref{eq:Tmatrix}), we need 2a) a way to find $G_0(R=0;\omega)$ and 2b) a way to find
the correct phases of $G_0(R;\omega)$. Once this is done, we Fourier transform
to get $\tilde{G}_0(k;\omega)$ and, thence, $\eta(k;\omega)$. We show results of this method for various cases
of dispersion and lifetimes. Then, we discuss what kind of data sets are desirable and how the method extends
to the superconducting case.

\begin{figure}
    \resizebox{120mm}{!}{\includegraphics{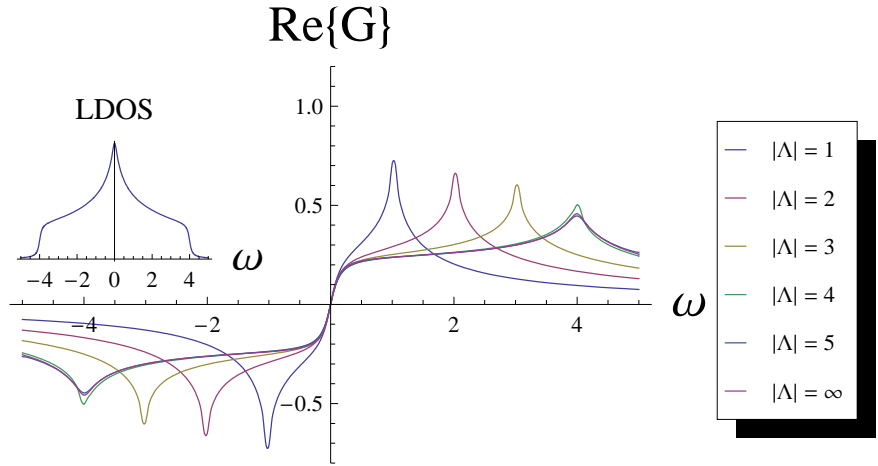}}
    \caption{In this figure, we demonstrate the effect of Kramers-Kroning to an example LDOS where we limit the integral
by a finite cut-off, $Re[G(r,r;\omega)] = P\int^{\Lambda}_{-\Lambda} \frac{n(r,r;x)}{(\omega - x)}$. The example LDOS
(see inset) is for a nearest-neighbour hopping model at half-filling, 
$\tilde{G}_0(k;\omega) = (\omega - i 0.1t + 2 t (Cos[k_x]+Cos[k_y]))^{-1}$ and $t=1$. Around the Fermi energy($\omega=0$ in this
case), we see that even for $|\Lambda| = 3$, $Re[G]$ agrees well upto around $|\omega|=1$. One can quantitatively show
that this error is at most $Log|\frac{\Lambda+\omega}{\Lambda-\omega}| \approx 2|\omega/\Lambda|$ in units of $n$
and we do much better than that(see Supplementary).}
    \label{fig:KK}
\end{figure}

The first step of the analysis method is to convert the LDOS data to $G(r,r;\omega)$. This will be achieved
through a Kramers-Kronig relation the propagator satisfies,  $Re[G(r,r;\omega)] = P\int n(r,r;x)/(\omega - x)$ 
where the principle value integral is over the 
real line. Since the LDOS is nonzero only within a finite bandwidth \cite{spectralweight}, this integral is 
over a finite range of energies. 
In general, in a real experiment one might have information only over part of the bandwidth 
in which case, we can definitively apply this method only
to an energy range that is well within the measured energy range, where even the incomplete
spectrum can be fruitfully used as demonstrated in
Fig. \ref{fig:KK}. This is very often the/one of the 
interesting energy ranges(e.g. around the Fermi energy for metals or the nodal energy for cuprates). 
We can also apply some form of extrapolation to construct LDOS data over the full bandwidth
\cite{extrapolation}. Kramers-Kronig has been 
applied successfully to other spectroscopies, e.g. Electron Microscopy (see \cite{KramKron}), thus giving us reason
that they be applied to STM data as well.

\begin{figure}
\resizebox{100mm}{!}{\includegraphics{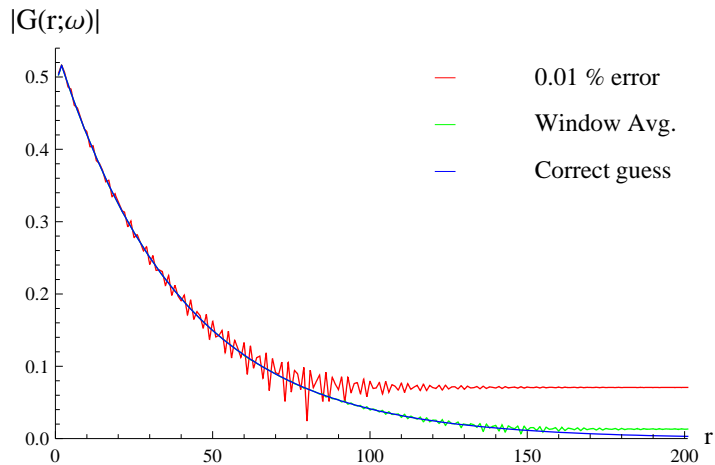}}
\caption{Above is shown $|G_0(R;\omega)|$ on a 200-site window around an impurity extracted with various start guesses for
$G_0(0;\omega)$.}
\label{fig:Guess}
\end{figure}

We now discuss the second step : how
to invert Eq. (\ref{eq:Tmatrix}) at a fixed energy. We are only concerned with $r=r'$.
We set $r_{imp}=0$. The first step is to find out the first term on the right hand side
of Eq. (\ref{eq:Tmatrix}), $G_0(R=0;\omega)$. This will be done through a minimization procedure
where a cost function would penalize incorrect guesses for $G_0(0;\omega)$. 
Given a $G_0(0;\omega)$ guess(which is independent of $R$ 
if the free propagator is that of a translationally 
invariant system), we can solve for $T(\omega)$ by solving Eq. (\ref{eq:Tmatrix}) for $r=r'=r_{imp}$
(Furthermore, we can calculate the impurity strength $U$ from $T(\omega)$). Once $T(\omega)$ is
known, we can solve for $G_0(R;\omega)$ as
\begin{equation}
G_0(R;\omega) = \sqrt{\frac{G(r,r;\omega)-G_0(0;\omega)}{T(\omega)}} 
\label{eq:invertT}
\end{equation}
In Green's function theory, one can show that the magnitude $|G_0(R;\omega)|$ monotonically
decays to zero for large $R$ (exponentially in $R$ in one dimensions and as square root of $R$ in two dimensions,
see Supplementary) for dispersion that have convex energy contours.
We demonstrate this effect in 1D and also show the effect of incorrect $G_0(0;\omega)$ on 
extracted $|G_0(R;\omega)|$ in Fig. \ref{fig:Guess}.
We see how an incorrect guess for $G_0(0;\omega)$ spoils the monotonic decay of $|G_0(R;\omega)|$.  
The reason for the deviation from monotonicity is as follows : Given our (incorrect) guess of $|G_0(0;\omega)|$,
we can decompose the incorrect $G_0(R;\omega)$ as $G_0^{correct}(R;\omega)+ G_0^{error}$ where $G_0^{error}$
is a constant. Therefore,
$|G_0(R;\omega ) |  = |G_0^{correct}(R;\omega ) | +|G_0^{error}|
+2|G_0^{correct}(R;\omega) | |G_0^{error}| \times
Cos(Arg[G_0^{correct}(R;\omega)]-Arg[G_0^{error}])$,
and it is the final cosine term in the above expression which spoils the monotonicity even 
for large $R$. Moreover, the $|G_0^{error}|$ term would also not let the Green's function decay to
zero as $r \rightarrow \infty$. This
motivates a minimization using a cost function that penalizes deviation from the smooth
decay of extracted $|G_0(R;\omega)|$ for finding the correct $|G_0(0;\omega)|$ \cite{cost}. A good start guess
for $G_0(0;\omega)$ is to take a spatial average of $G(r,r;\omega)$ over the whole data set
around the impurity. One can show that the error in the guessed $G_0(0;\omega)$ is $1/L^2$($1/L^d$ in $d$ dimensions, see Supplementary) 
suppressed compared  to the guessed $G_0(0;\omega)$, and if the window were infinite, the spatial average 
of $G(r,r;\omega)$ would exactly equal $G_0(0;\omega)$.

With the correct $G_0(0;\omega)$, we still get $G_0(R;\omega)$ only up to a phase of $\pi$.
Capturing this phase is crucial to get the
correct $\tilde{G}_0(k;\omega)$ upon Fourier transforming. To get the correct phase,
we start with the observation that the phases have to be smooth and well-behaved as a function of $R$
because $G_0(R;\omega)$ is differentiable with respect to $R$ \cite{imp}.
We use this property to fix the phase of the square of $G_0(R;\omega)$, i.e. we select that branch of
the argument function when evaluating the phase of $G_0(R;\omega)^2$ which maintain the aforesaid smoothness. 
We start by making a spatial list of 
the phases as given by the $Arg(z)$ function which restricts the phase obtained to one branch of the Argument
function. Then,
we start at $R = 0$. As we move away from the origin, we multiply phase factors of $e^{i 2 m \pi}$ to
$G_0(R;\omega)^2 = |G_0(R;\omega)^2| e^{i \phi_{principal}}$ for all $R$, the 
$m$'s being so chosen that if $|R'| > |R|$ then $\phi'_{principal} + 2 \pi m' > \phi_{principal} + 2 \pi m $.
Once that is done, the phase of $G_0(R;\omega)$
is just half that of $G_0(R;\omega)^2$. We demonstrate the working of this phase reconstruction 
method in Fig. \ref{fig:Phase} (see Supplementary for a flowchart of the method).

\begin{figure}
    \begin{tabular}{ll}
      a) \resizebox{60mm}{!}{\includegraphics{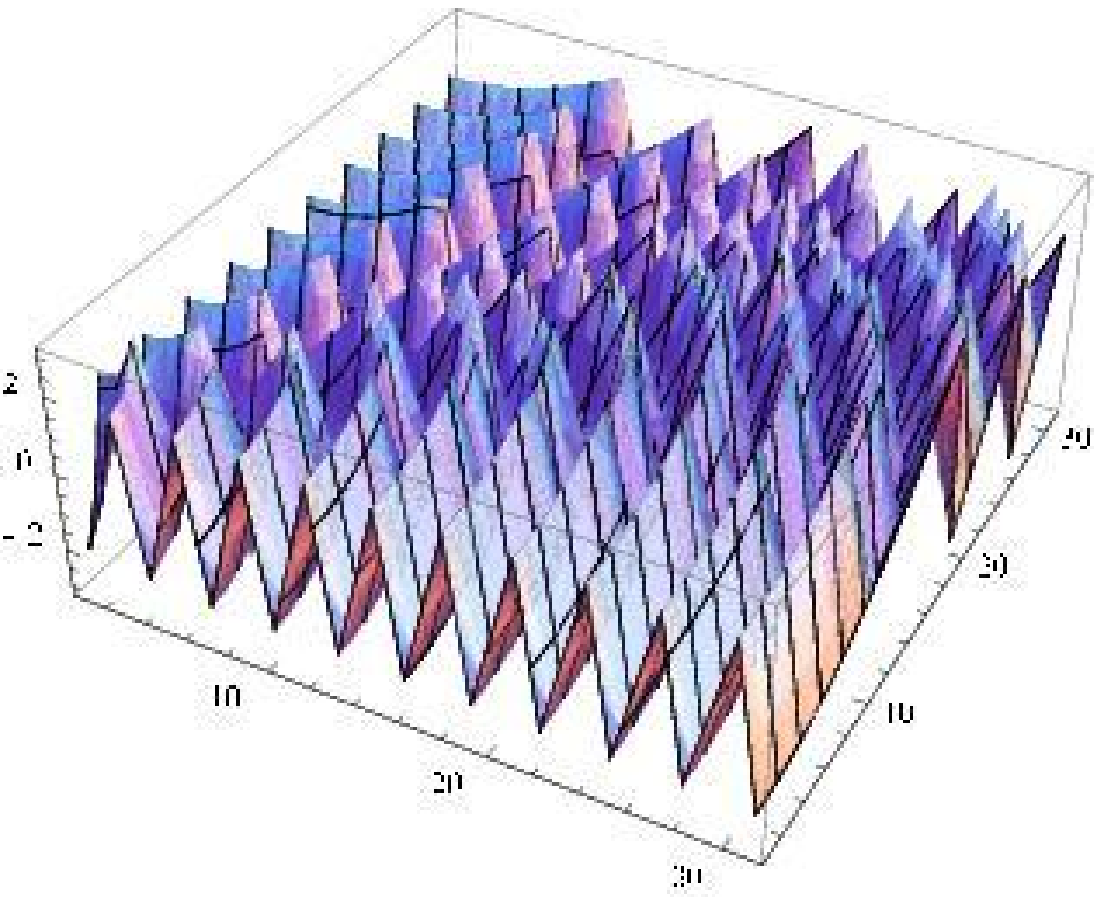}} & b) \resizebox{60mm}{!}{\includegraphics{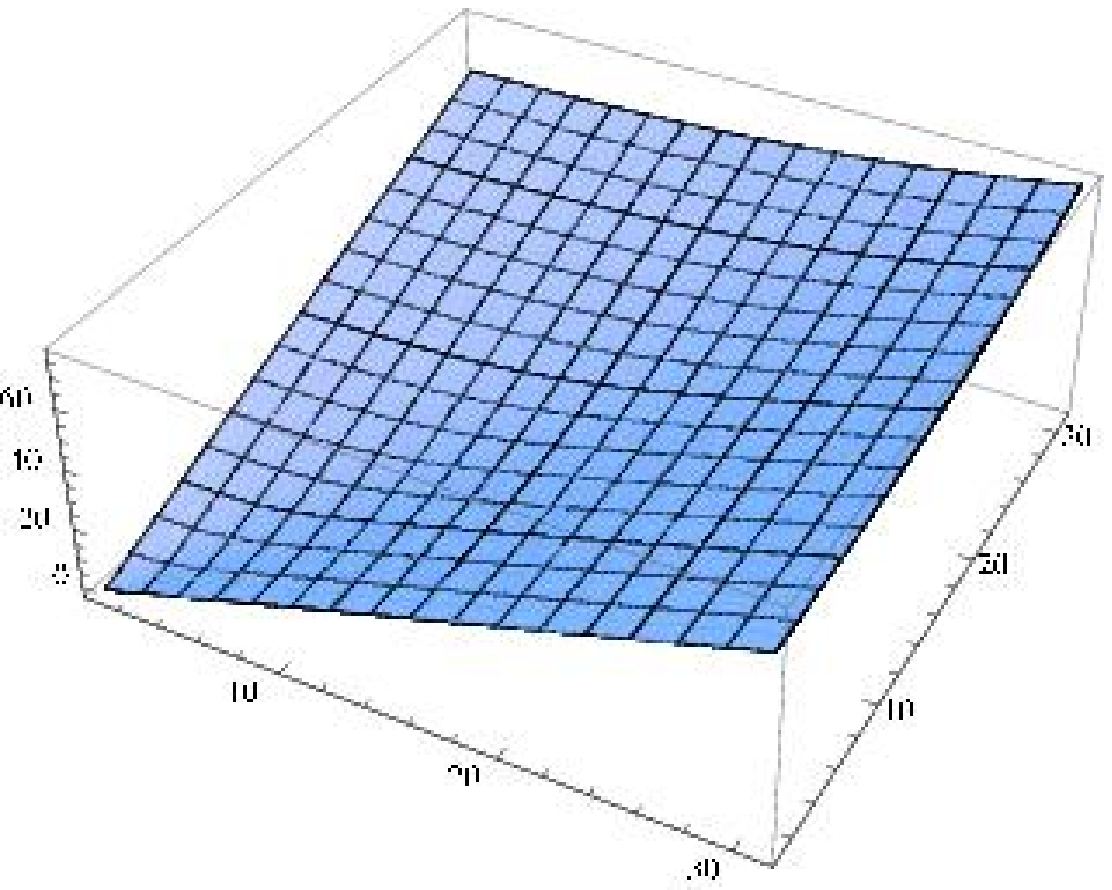}} \\
      c) \resizebox{60mm}{!}{\includegraphics{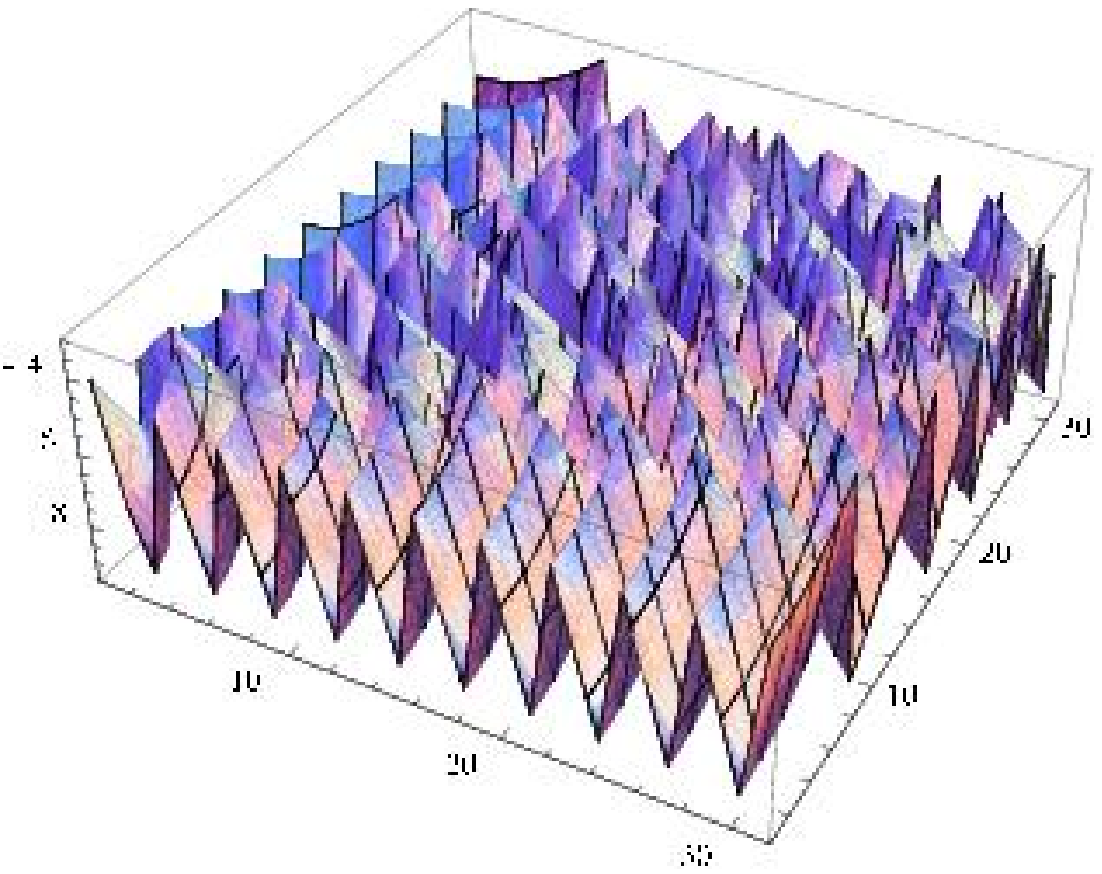}} & d) \resizebox{60mm}{!}{\includegraphics{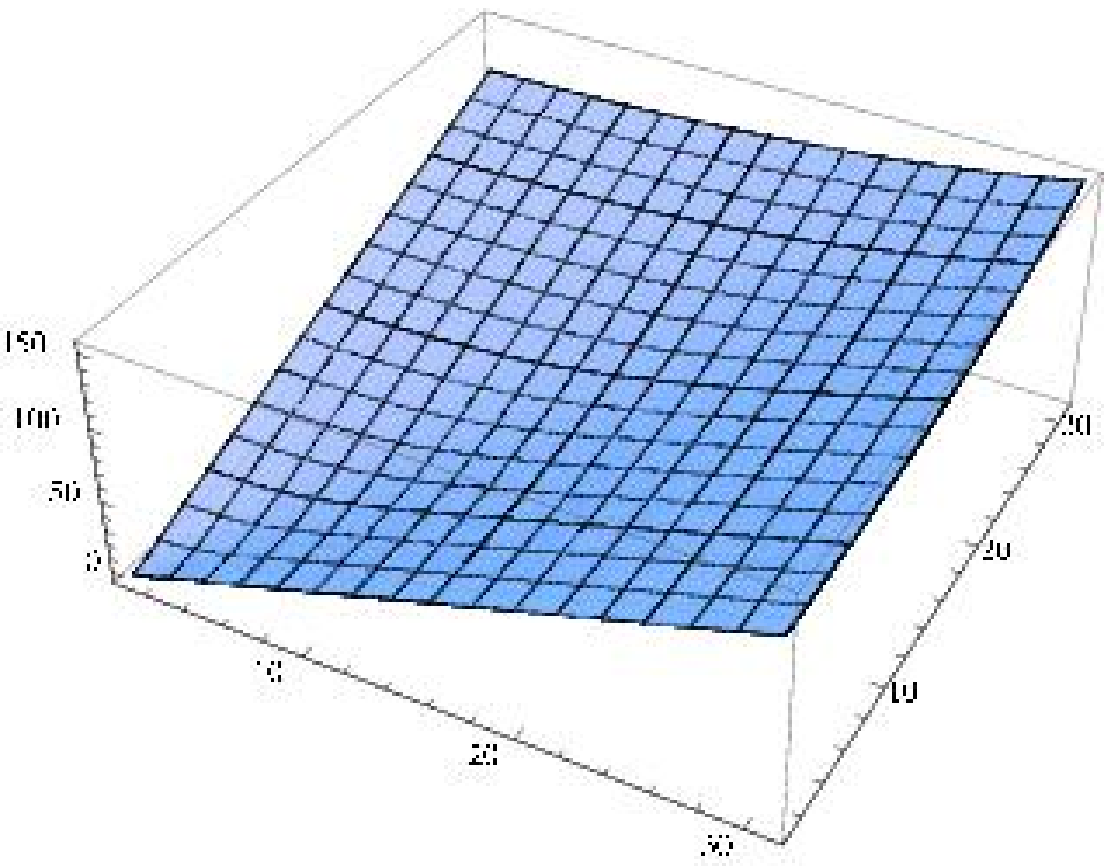}} \\ 
    \end{tabular}
    \caption{In the panels above we show the phases of $G_0(R;\omega)$ (a) and b)) and $G_0(R;\omega)^2$ (c) and d)) 
in the first quadrant of size 30x30 lattice sites 
around an impurity at a fixed $\omega$($=-t$). a) and c) show the phases as evaluated by the Arg(z) function restricted
to one branch. b) and d) show the smooth phases as reconstructed using the reconstruction algorithm. The ratio of phases in
b) and d) is identically two over the whole quadrant, even though the ratio of phases in a) and c) does not behave in such
a regular manner. $\epsilon(k) = - 2 t (Cos[k_x]+Cos[k_y])$ and $\eta(k,\omega=-t)= 0.1 t$ in this example.}
    \label{fig:Phase}
\end{figure}

\begin{figure}
    \begin{tabular}{ll}
      a) \resizebox{70mm}{!}{\includegraphics{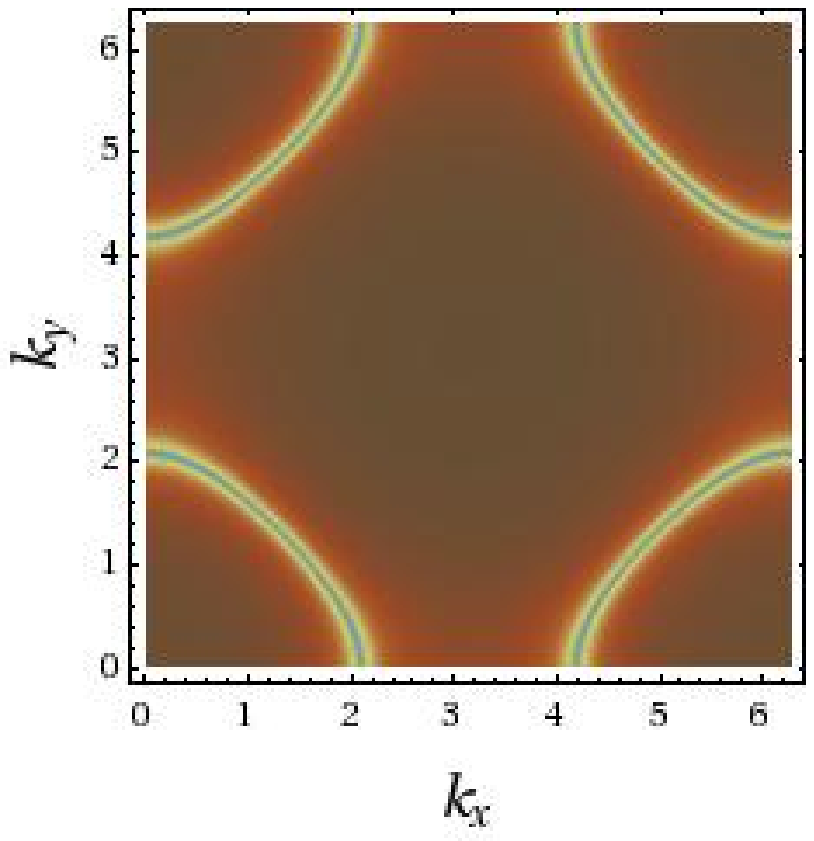}} & b) \resizebox{70mm}{!}{\includegraphics{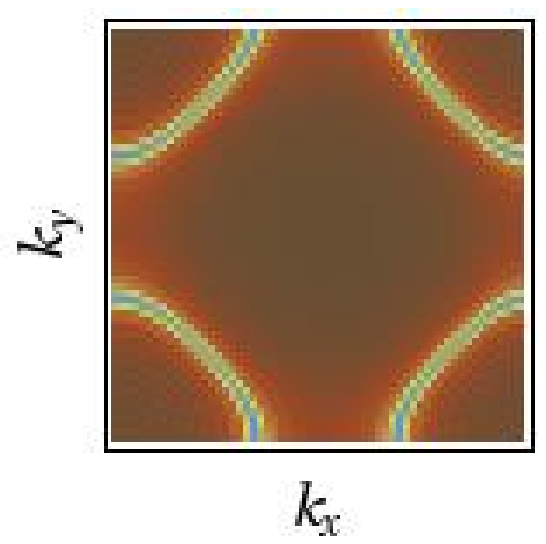}} \\
      c) \resizebox{70mm}{!}{\includegraphics{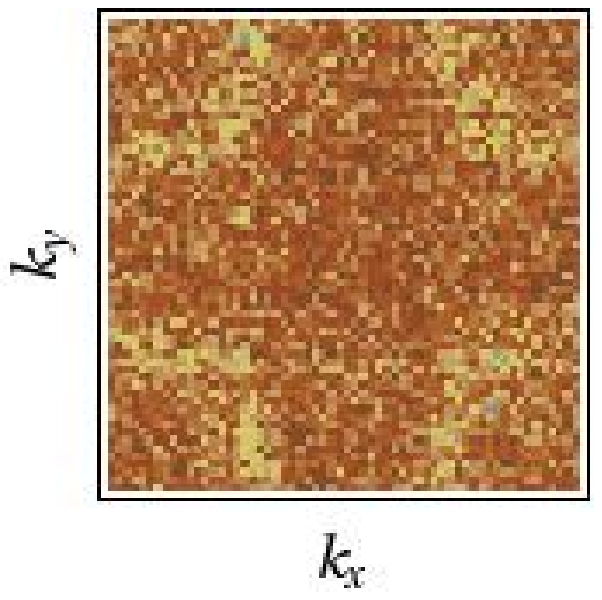}} & d) \resizebox{70mm}{!}{\includegraphics{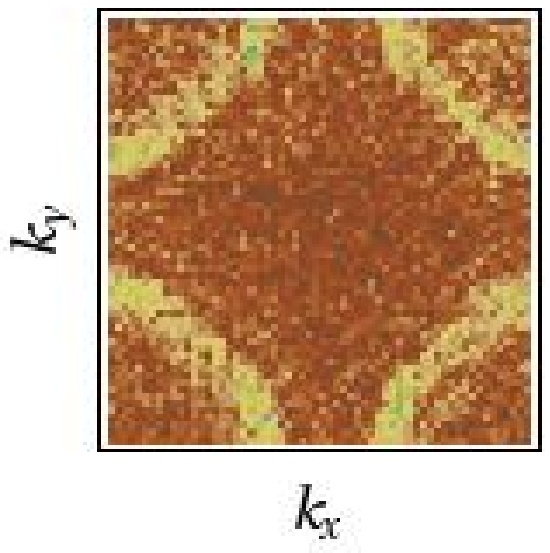}} 
    \end{tabular}
    \begin{tabular}{lll}
      e) \resizebox{40mm}{!}{\includegraphics{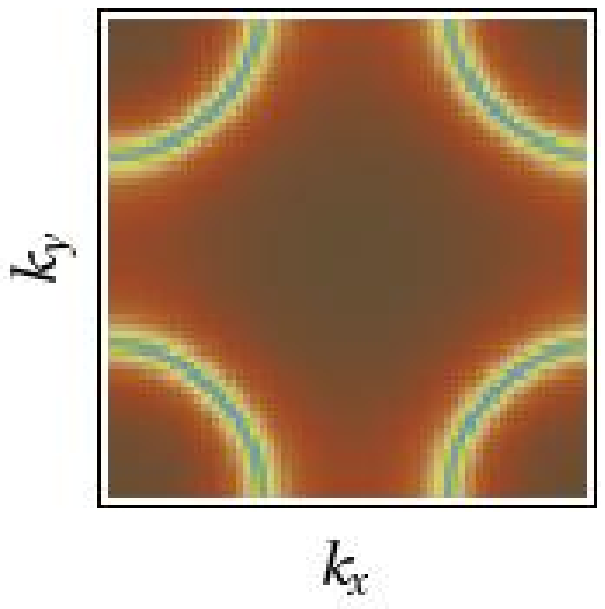}} & f) \resizebox{40mm}{!}{\includegraphics{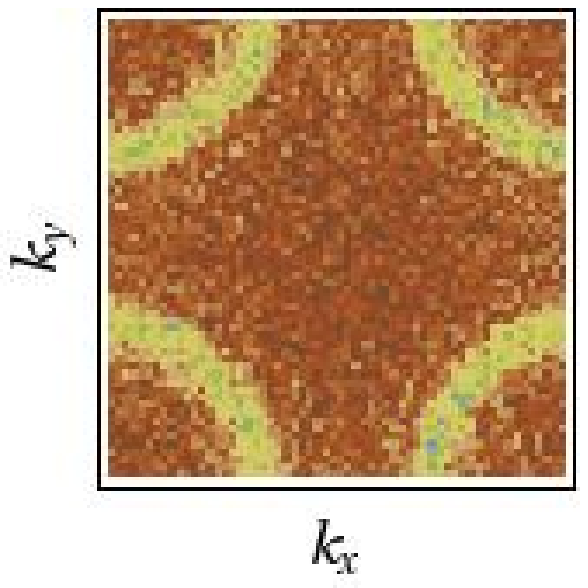}}
		& g) \resizebox{40mm}{!}{\includegraphics{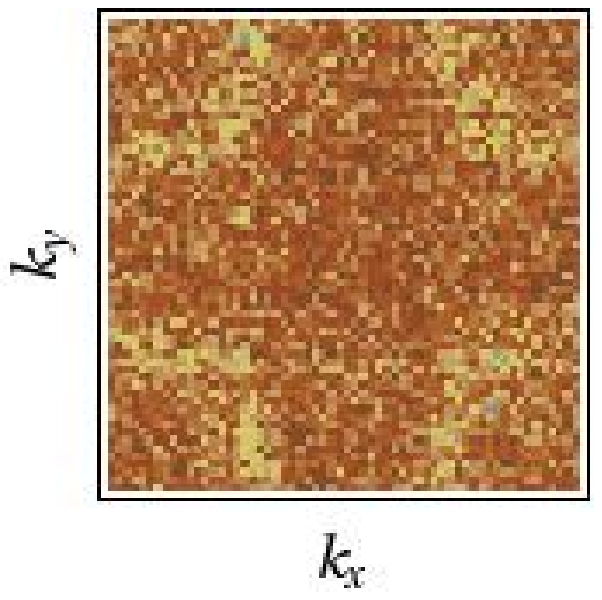}} \\
    \end{tabular}
    \caption{In these figures we are plotting $|(\omega + i \eta(k,\omega) - \epsilon(k))^{-1}|$ as a function of $\vec{k}$ over
a Brillouin Zone $(0,2\pi)\times(0,2\pi)$ at a 
fixed $\omega$($=-t$ for the above plots). In a) we show the input form resulting out of our choice of input for $\eta$,
where $\epsilon(k) = - 2 t (Cos[k_x]+Cos[k_y])$ (nearest-neighbour hopping) and $\eta(k,\omega=-t)= 0.1 t$; in b) we show
the form extracted using the proposed analysis scheme when no noise was added to the STM data calculated numerically. One sees the 
limitation in momentum resolution in the form of ``blockiness" introduced by having a finite window. This ``blocky" momentum resolution 
gets better or worse with greater or smaller window sizes.
In c) and d) we show the results of the analysis scheme to data
with $1 \%$ and $0.05 \%$ Gaussian errors added respectively. We have done similar analyses 
for different energy values and different
forms of $\eta$ and in e), f) and g), we show the corresponding results for $\epsilon(k) = - 2 t (Cos[k_x]+Cos[k_y]) 
- 4 (0.2 t) (Cos[k_x]*Cos[k_y])$ (nearest and second-nearest neighbour hopping)
and $\eta(k,\omega=-t)= 0.25 t + 0.1 t (Cos[k_x]+Cos[k_y])$ as another example. }
    \label{fig:Results}
\end{figure}

With the correct phases, we are now ready to Fourier transform the extracted $G_0(R;\omega)$ to get
$\tilde{G}_0(k;\omega)$ and $\eta(k;\omega)$ with our knowledge of $\epsilon(k)$.
Moreover, the extracted $\tilde{G}_0(k;\omega)$ also has to satisfy the exclusive momentum dependence 
of $\omega - Re[\tilde{G}_0(k;\omega)^{-1}]$. In Fig. \ref{fig:Results}, we show how this method performs with and without
error and we see that it performs well for error magnitudes less than 0.25 $\%$. 
For the panels Fig. \ref{fig:Results} a-d, the form of $\eta$ had no momentum dependence, and this kind of fitting 
form has been proposed in \cite{Jacob} for Cuprates and has been theoretically discussed
in \cite{Scalapino}. In general, we expect the lifetime function to have few (low) harmonics of $k$ similar
to the dispersion. Thus, our analysis scheme would serve the purpose of finding the most general $\eta(k;\omega)$ that
is consistent with STM data. We can extract an approximate analytic form for $\eta$ by doing a least-squares fit of
the extracted $\eta$ to a function of $k$ containing a few harmonics in the Brillouin zone. 
The approximate analytic form can then be compared to theoretical proposals.

\begin{figure}
    \begin{tabular}{ll}
      a) \resizebox{80mm}{!}{\includegraphics{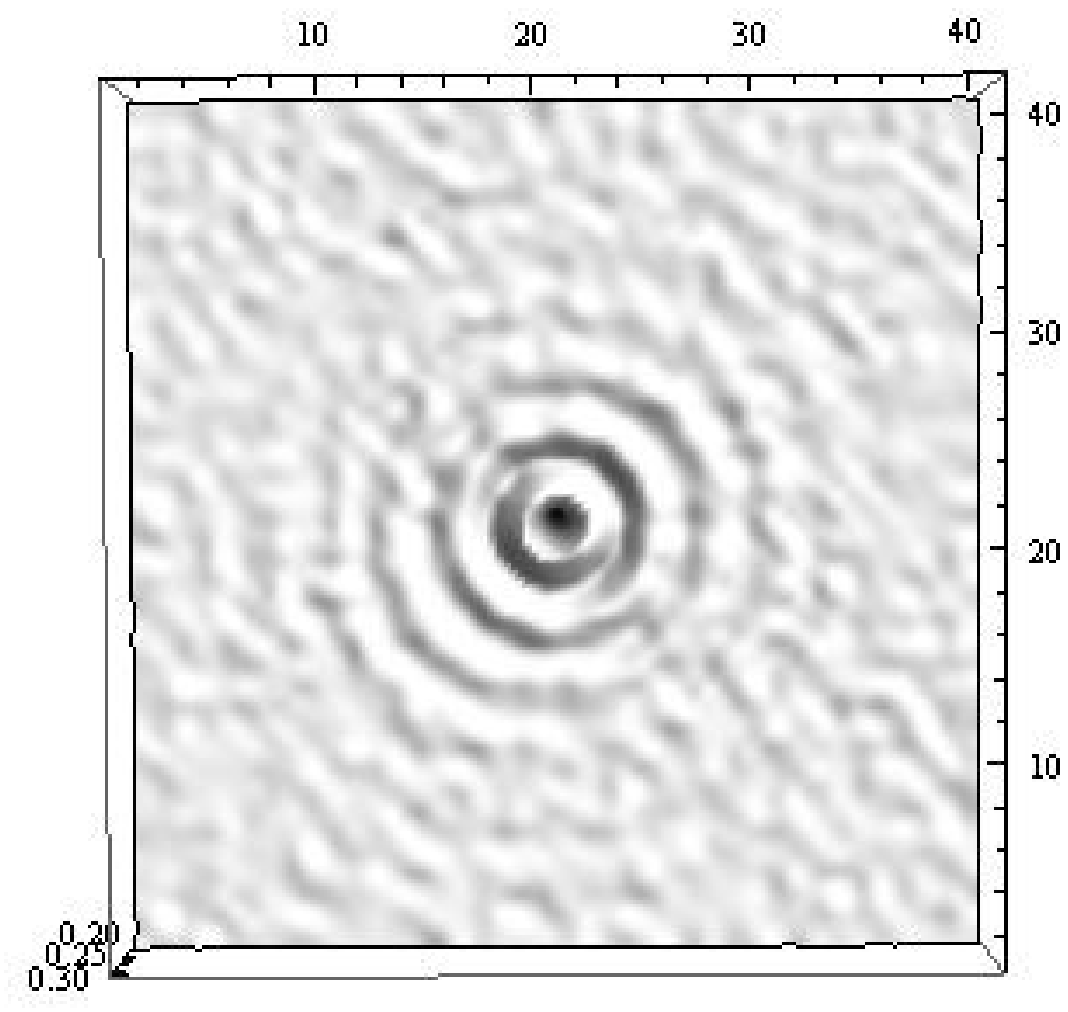}} & b) \resizebox{60mm}{!}{\includegraphics{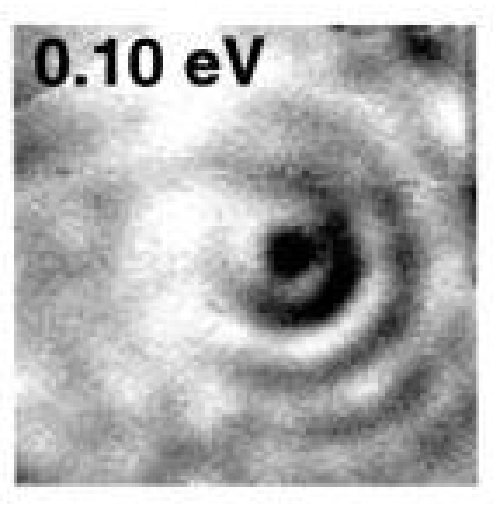}} \\
     \end{tabular}
        \caption{In this figure we compare an experimental data set on InAs surface(taken from \protect\cite{semiconddata} with
due permission from APS and the authors) with
        a numerically calculated LDOS data set with error 0.5 $\%$ added. This figure serves to illustrate that there exist
        data sets, perhaps within operable error range, to which the scheme can potentially be applied.}
    \label{fig:dataset}
\end{figure}

At this point, we comment
on what kind of data sets would be ideal for such an analysis. In Fig. \ref{fig:dataset}, we show an example of data set
seen in a real experiment. We show how it is similar to a theoretical data set(calculated numerically) which has a
lifetime broadening. Thus, we would expect that if we observe a few of the "Friedel oscillation"-like
rings around the point impurity, this analysis scheme should work. Moreover, if FT-STS applied to a single point
impurity data shows reliable QPI peaks, then we believe that the data set would have good enough spatial resolution
to resolve the momentum dependence of lifetime $\eta$ to the same momentum resolution as that of the FT-STS results.
We can improve on this by taking an average over data sets around multiple point impurities to improve signal
to noise. 
Now, we will sketch how this method of analysis can be extended to superconducting case using d-wave
superconductors(pertinent to Cuprate phenomenology) as our example. In Nambu's
two component notation, the free superconducting propagator looks like
\begin{equation}
\tilde{G}_0(k;\omega)^{-1} =
\big( \begin{array}{lr}
			\omega - i \eta_{e}(k,\omega) - \epsilon(k) &  \Delta(k)        \\
			\Delta(k)*    & \omega - i \eta_{h}(k,\omega) + \epsilon(k)  \\
		\end{array} \big)
\label{eq:freeSCprop}
\end{equation}
where $\epsilon(k)$ is the bare dispersion and $\Delta(k)$ is the (d-wave) gap of the cuprate in
question. These are assumed to be known(through other experiments). As before, we want to determine 
the electron/hole lifetime. 
The first simplification is the relation $\eta_h(k,\omega) = -\eta_e(k,-\omega)$. The proof of
this relation in outlined in the supplementary information to this manuscript and it follows
by showing $\Sigma_{22}(- \omega - i \delta) = - \Sigma_{11}(\omega + i \delta)$. 
This relation implies $G_0(R,\omega)_{22} = - G_0(R,-\omega)_{11}$
and  $G_0(R,\omega)_{12} = G_0(R,-\omega)_{12}$.
Now, as before, we imagine there is a point impurity which result in a two-component
T-matrix. One can show that this T-matrix has no off-diagonal entries(for an ordinary potential
impurity) since $G_0(R=0,\omega)_{12} = 0$ owing to the d-wave symmetry of the gap function.
One can further show that $T_{22}(-\omega) = - T_{11}(\omega)$ and, resultantly,  $G(r,r;\omega)_{22} = - G(r,r;-\omega)_{11}$.
For $r=r_{imp}$, we have $G_{11} = G_{0,11} + T_{11} G_{0,11}^2$ and $G_{22} = G_{0,22} + T_{22} G_{0,22}^2$.
Using $G(r,r;,\omega)_{22} = - G(r,r;,-\omega)_{11}$, we can thus determine $T_{11}$ and $T_{22}$
given a guess for $G_0(R=0,\omega)_{11}$ (which will again be determined by demanding the
monotonicity of $G_0(R,\omega)_{11}$ ). For $r \neq r_{imp}$, we have 
\begin{eqnarray}
G_{11}(r,r;\omega) &
 = & G_0(0,\omega)_{11} + T_{11}(\omega) G_0(r-r_{imp},\omega)_{11}^2  \\
& & + T_{22}(\omega)  G_0(r-r_{imp},\omega)_{12} G_0(r-r_{imp},\omega)_{21} \nonumber \\
G_{22}(r,r;\omega) &
 = & G_0(0,\omega)_{22} + T_{22}(\omega)  G_0(r-r_{imp},\omega)_{22}^2 \\
& & + T_{11}(\omega)  G_0(r-r_{imp},\omega)_{12} \nonumber
G_0(r-r_{imp},\omega)_{21}.
\end{eqnarray}
Again using $G(r,\omega)_{22} = - G(r,-\omega)_{11}$, now with the knowledge of $T_{11}$ and $T_{22}$,
we can solve the above equations for $G_0(R,\omega)$ upto a phase of $\pi$ (which we reconstruct as before) 
at each $\omega$ for all $r$ in the dataset, following which we Fourier transform to 
extract $\tilde{G}_0(k,\omega$ and $\eta_e(k,\omega)$.

In conclusion, we demonstrated an analysis scheme which holds promise to extract
lifetimes from STM data in various systems ranging from metals and semiconductors
to strongly correlated compounds to superconductors. Some final remarks are in order. 
We demonstrated the proposed analysis scheme in case of a point impurity, but it can be extended to the case of an extended
impurity too. The resulting complication will be that now we would have to guess more than just $G_0(R=0;\omega)$
(e.g. if the impurity extends over two adjoining sites $r_1$ and $r_2$, then $G(r,r;\omega)$ will
be a function of $G_0(0;\omega)=G_0(r_1,r_1;\omega)=G_0(r_2;r_2;\omega)$ and $G_0(1;\omega)=G_0(r1,r2;\omega)$).
This scheme is inherently local, where
we would be analyzing data around a single impurity. Thus, it would really utilize the local information
that STM affords us with. There have been other examples of data analysis done on STM data previously to extract localinformation(\cite{Cobalt},\cite{Hudson}). 
In this sense, we would do better than ARPES where the signal
is averaged over an area of the sample equal to the beam size, iff the STM experiment has good signal to noise.
Similarly, we can overcome inhomogeneity issues for dirty systems, in which case we would concentrate this analysis on
a homogeneous patch similar in spirit to Hudson's analysis \cite{Hudson} and to a previous work \cite{Echoes}.
\\
\textbf{Acknowledgements} : This work has been supported by the NSF grant DMR-0552461. I acknowledge the extremely
valuable suggestions of C. L. Henley, especially regarding the checks on robustness of the method to error
and phase fixing, and a critical reading of the manuscript. I also thank him and H. J. Changlani, Milan Allan and 
J. C. Davis for helpful discussions.

\section{Appendix : Supplementary Information}

\subsection{Limit on the Error introduced by Kramers-Kroning Integration}
The Kramers-Kronig relation relates the real part of a Green's function to the
imaginary part as follows
\begin{equation}
Re[G(r,r;\omega)] = - \frac{1}{\pi} P \int^{\infty}_{-\infty} dx \frac{Im[G(r,r;x)]}{(\omega - x)} =
P \int^{\infty}_{-\infty} dx \frac{n(r;x)}{(\omega - x)}.
\label{eq:KK}
\end{equation}
If we limit the integral by cut-offs $\Lambda_+$ and $\Lambda_-$, then the error $E$ introduced is
\begin{equation}
P \int^{\Lambda_-}_{-\infty} dx \frac{n(r;x)}{(\omega - x)}+P \int^{\infty}_{\Lambda_+} dx \frac{n(r;x)}{(\omega - x)}.
\end{equation}
Even if we were to make the really bad approximation that $n(r;x) = n(r;\omega)$ for all $x$ (and this is a really
bad approximation, since $n(r;x)$ decays to zero as $x \rightarrow \infty$), we get
\begin{equation}
E < n(r;\omega) Log \frac{|\omega - \Lambda_-|}{|\omega - \Lambda_+|}
\end{equation}
since $n(r;x) > 0$ for all $x$. Thus, if $\omega/\Lambda_- << 1$ and $\omega/\Lambda_+ << 1$, then the error in 
$Re[G(r,r;\omega)]$ is less than $n(r,\omega)*(\frac{\omega}{\Lambda_+} + \frac{\omega}{\Lambda_-} + Log\frac{|\Lambda_-|}{|\Lambda_+|})	$.
Since, $Re[G(r,r;\omega)]$ and $n(r;\omega)$ carry the same dimensions and $n(r;\omega) < 1$, this is at most  an
 $O(\frac{|\omega|}{|\Lambda|})$ error.

\subsection{Proof of  Monotonic decay of $G_0(\RR;\En)$ for large $\RR$}

In two dimensions, $G_0(\vec{R})$ on a lattice is given by the formula
\begin{eqnarray}
\label{eq:contal1}
G_0(\R,\R_{imp};\En) & \equiv & G_0(\R-\R_{imp};\En) = G_0(\RR;\En) \nonumber \\
&=& \lim_{\delta \rightarrow 0^+} \frac{1}{(2\pi)^2} \int_{B.Z.} d \K \frac{e^{i \K. \RR}}{\En+i\delta-\epsilon(\K)} \nonumber \\
&=& \frac{1}{(2\pi)^2} \int^\pi_{-\pi} dk_x e^{i k_x R_x} \int^\pi_{-\pi} dk_y \frac{e^{i k_y R_y}}{\En + i\delta -\epsilon(\K)}
\end{eqnarray}
Let us look at the $k_y$ integral for a particular $k_x$. The denominator vanishes for certain values of $k_y$ thus motivating 
the conversion of the $k_y$ integral to a contour integral. The mapping $z=e^{i k_y}$ achieves the conversion which also maps
the integral from $-\pi$ to $\pi$ to a contour integral over the unit circle. The periodicity of the integrand over the zone
ensures the analyticity of the resulting complex integrand. Thus,
\begin{eqnarray}
\label{eq:contal2}
G_0(\RR;\En) &=& \frac{1}{(2\pi)^2} \int^\pi_{-\pi} dk_x e^{i k_x R_x} \oint_{U. C.} \frac{dz}{i z} \frac{z^{R_y}}{\En + i\delta-\epsilon(k_x,z)} 
\end{eqnarray}
For a particular $\En$ energy contour and $k_x$, we get two poles (See Fig.). Expanding the denominator around the 
poles gives us
$\En + i \delta - \epsilon(k_x,z) = - \frac{\hbar v_{g_y} (\En, k_x)}{i z_p} (z - z_p(1-\frac{\delta}{\hbar v_{g_y} (\En, k_x)}))$.
The poles $z_p$s are defined by $\epsilon(k_x,z_p)=\En$ and $\hbar v_{g_y} (\En, k_x) \equiv \frac{\partial \epsilon(\K)}{\partial k_y}$
is the group velocity along $y$ direction. 
We need only worry about the $(z-z_p)$ term in the expansion of the denominator since other expansion terms will yield zero residues.
From the $(1-\frac{\delta}{\hbar v_{g_y} (\En, k_x)})$ factor in the expansion, we realize
that the pole where the sign of the $\hbar v_{g_y}$ is same as the positive $\delta$ will be ``pulled" inside the unit circle while
the other pole will be ``pushed" out of the unit circle. Thus, when we do $k_x$ integral, only one half of the $\En$ \emph{energy}
contour (not to be confused with the complex contour; to distinguish we will call $\En$ contours as energy contours) will contribute to
the integral. In the process, we have converted the 2D integral over the zone into an integral over part of the energy contour.
Filling in the steps,
\begin{eqnarray}
\label{eq:contal3}
G_0(\RR;\En) &=& \frac{1}{(2\pi)^2} \int^\pi_{-\pi} dk_x e^{i k_x R_x} \oint_{U. C.} \frac{dz}{i z} \frac{z^{R_y}}{-\frac{\hbar v_{g_y}
                 (\En, k_x)}{i z_p} (z - z_p(1-\frac{\delta}{\hbar v_{g_y} (\En, k_x)}))} \nonumber \\
&=&-\frac{1}{(2\pi)^2} \int^\pi_{-\pi} dk_x e^{i k_x R_x} 2 \pi i  \frac{z^{R_y}_p}{\hbar v_{g_y} (\En, k_x)} \nonumber \\
&=& -\frac{1}{(2\pi)^2}2 \pi i \int d k_x \frac{e^{i k_x R_x} e^{i k_{y_p}(\En, k_x) R_y}}{\hbar v_{g_y} (\En, k_x)} \nonumber \\
&=& \frac{1}{2\pi i}  \oint_{\mathrm{sgn(\delta)=sgn(v_{g_y} (\En, s))}} ds \frac{e^{i \K(s,\En).\RR}}{|\nabla \epsilon(\K(s,\En))|}
\end{eqnarray} 
where the last step was achieved by converting the element $d k_x$ to a parameter $s$ characterising the $\En$ energy contour
and we integrate over that part of the contour where the sign of $\delta$ is same as $v_{g_y}$.

For large $\RR$, i.e. far from impurity, we notice that the phase $e^{i \K(s,\En).\RR}$ varies rapidly and thus stationary
phase approximation can be applied. The phase factor is stationary at points on the energy contour where the group velocity is 
along the $\RR$ direction since $d(\K(s,\En).\RR)/ds=\RR.d\K(s,\En)/ds=0$ only when $\RR$ is perpendicular to $d\K(s,\En)/ds$ 
and $d\K(s,\En)/ds$, being the tangent to the energy contour, is perpendicular to the group velocity. Therefore,
\begin{eqnarray}
\label{eq:statphase}
G_0(\RR;\En) &=& \frac{e^{i \pi/4}}{2\pi i}\frac{1}{|\nabla \epsilon(\K_{dom}(\RR,\En))|} 
\sqrt{\frac{2 \pi}{|\RR| |d^2 \K(s,\En)/ds^2|_{\K_{dom}(\RR,\En)}}} e^{i\K_{dom}(\RR,\En)\cdot\RR} \nonumber \\
\end{eqnarray}
where $\K_{dom}(\RR,\En)$ is the $\K$ corresponding to which the group velocity at energy $\En$ is along $\RR$ and, thus, is a
function of $\RR$ (only through $\hat{R}$) and $\En$. For a convex dispersion function $\epsilon(\K)$, we will have only one
$\K_{dom}$ and thus
\begin{equation}
|G_0(\RR;\En)| \propto \frac{1}{\sqrt{|\RR|}} 
\end{equation}
for large $\RR$. This proves the monotonic decrease of $G_0(\RR;\En)$ when the lifetime is infinitesimal. When we have a
finite lifetime due to self-energy processes, the propagator in momentum space looks like 
$\tilde{G}_0(\K;\En)= (\En + i \delta - (\epsilon(\K) + i \eta(\K,\En)))^{-1}$ where the $\En$ 
might have undergone a chemical potential shift, and
the whole algebra in the above will go through similarly and we will get
\begin{equation}
|G_0(\RR;\En)| \propto \frac{1}{\sqrt{|\RR|}} e^{-\frac{\eta(\K_{dom}(\RR,\En))}{|\nabla \epsilon(\K_{dom}(\RR,\En))|}|\RR|}
\end{equation}
In one dimension, we only get the monotonic exponential decay for large $\RR$.

\section{Implementation of Cost Function for finding $G_0(R=0;\omega)$}
As mentioned in the main manuscript, the Cost function for a one-dimensional list 
of values for $|G_0(R;\omega)|$ (that is extracted given a guess $G_0(R=0;\omega)$) was 
\begin{equation}
Cost(\{z_r\}) = \displaystyle\sum\limits_{r} \frac{|z_{r+1}+z_{r-1}-2z_r|^2}{|(z_{r+1}+z_{r-1})/2|^2}
\end{equation}
where the list is $\{z_r\}$ and sum runs over all 3-tuples. We generalize this to two dimension
by evaluating the one-dimensional cost using the same formula for all one-dimensional slices
of the two-dimensional data set either along $x$ or $y$ direction. We do it this way because the two-dimensional
data set is symmetric with respect to interchanging $x$ and $y$ when there is no error. In the error-full
case, we can pre-process the data set to impose the symmetries of the square lattice. Thus the Cost
function is
\begin{equation}
Cost(\{z_{x,y}\}) = \displaystyle\sum\limits_{y} \displaystyle\sum\limits_{x} 
\frac{|z_{x+1,y}+z_{x-1,y}-2 z_{x,y}|^2}{|(z_{x+1,y}+z_{x-1,y})/2|^2}
\end{equation}

\begin{figure}[h]
\begin{tabular}{l}
a) \resizebox{160mm}{!}{\includegraphics{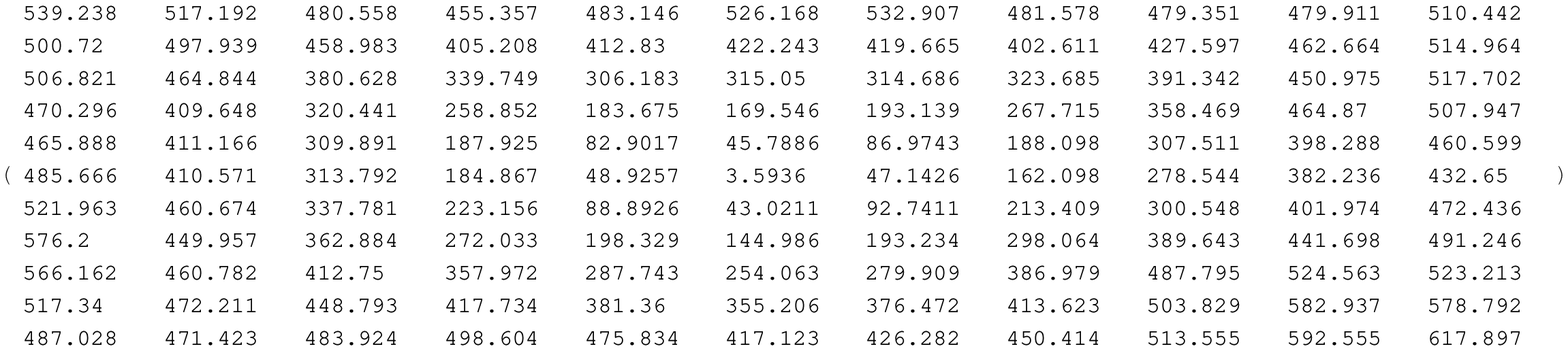}}  \\ \\
b) \resizebox{160mm}{!}{\includegraphics{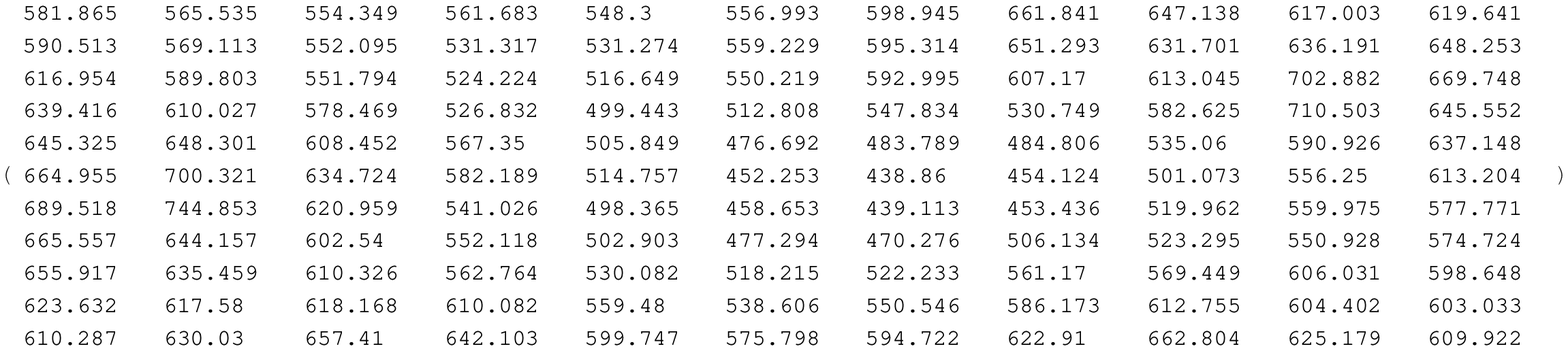}}\\ \\
c) \resizebox{160mm}{!}{\includegraphics{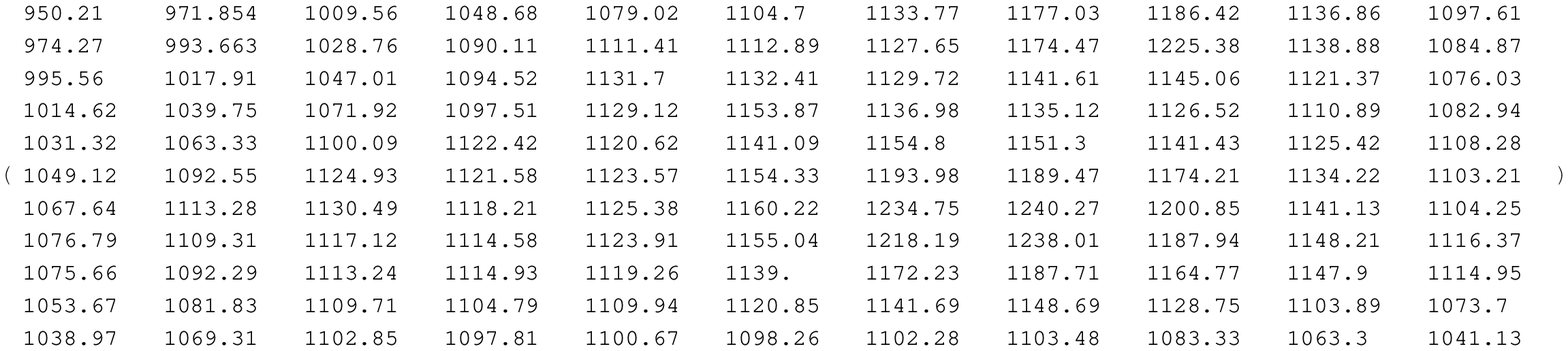}}
\end{tabular}
\caption{a) no-error, b) $0.1 \%$ error added and c) $0.5 \%$ error added to $G(r,r;\omega)$.
In this example, $\epsilon(\K) = -2 (Cos[k_x]
+Cos[k_y])$ and $\eta(\K) = 0.1$ and $\omega = -1$ in unit of $t$.}
\label{fig:Cost}
\end{figure}

We show the profile of this Cost function as a function of $G_0(R=0;\omega)$ guesses for 
the no-error case (which includes the numerical error incurred during two-dimensional Numerical
Integration in Mathematica) and error-full cases in Fig. \ref{fig:Cost}. We show it as a matrix where the center
point(6,6) corresponds to the correct $G_0(R=0;\omega)$ and the (7,7)-entry corresponds
to average over the $G(r,r;\omega)$ set (see next section). From point to point, we change the
guess by $Re[Avg(G(r,r;\omega))-G_0(R=0;\omega)]$ along $x$-direction and $Im[Avg(G(R;\omega))-G_0(R=0;\omega)]$
along $y$ direction.
We see that in the no-error case, the Cost function has minimum at the correct value of $G_0(R=0;\omega)]$.
In case of $0.1 \%$, it does well to within $Avg(G(r,r;\omega))-G_0(R=0;\omega)$. In case of $0.5 \%$,
it starts to seriously deviate and the best guess then would be $Avg(G(r,r;\omega))$.

\subsection{A Good Guess for $G_0(r,r;\omega)$}
Recalling that the T-matrix equation for scattering of point impurity is
\begin{equation}
G(r,r';\omega) = G_0(r,r';\omega) + G_0(r,r_{imp};\omega) \cdot T(\omega) \cdot G_0(r_{imp},r';\omega),
\label{eq:Tmatrix1}
\end{equation}
when we take the average of Eq. \ref{eq:Tmatrix1} with $r=r'$ over the window, the two terms on the right hand side 
average to(in two dimensions)
\begin{eqnarray}
\frac{1}{L^2} \sum_r G_0(0;\omega) & = & \big(\frac{2\pi}{L}\big)^2 \sum_k G_0(k;\omega) \\
\frac{1}{L^2} \sum_r G_0(r,0;\omega) T(\omega) G_0(0,r;\omega) & = & (\frac{2\pi}{L})^4 T(\omega) \sum_k G_0(k;\omega)^2 \\
\end{eqnarray}
Thus, we see that the second term is $1/L^2$($1/L^d$ in $d$ dimensions) suppressed compared to the 
first term, and if the
window were infinite, the spatial average of $G(r,r;\omega)$ would exactly equal $G_0(0;\omega)$.
For a finite but large enough window, it is a good guess for $G_0(0;\omega)$.

\subsection{Phase reconstruction algorithm}
\begin{figure}
    \begin{tabular}{ll}
      a) \resizebox{80mm}{!}{\includegraphics{inputphase.eps}} & b) \resizebox{80mm}{!}{\includegraphics{inputphaserecon.eps}} \\
      c) \resizebox{80mm}{!}{\includegraphics{inputphasesquare.eps}} & d) \resizebox{80mm}{!}{\includegraphics{inputphasesquarerecon.eps}} \\
	  e) \resizebox{80mm}{!}{\includegraphics{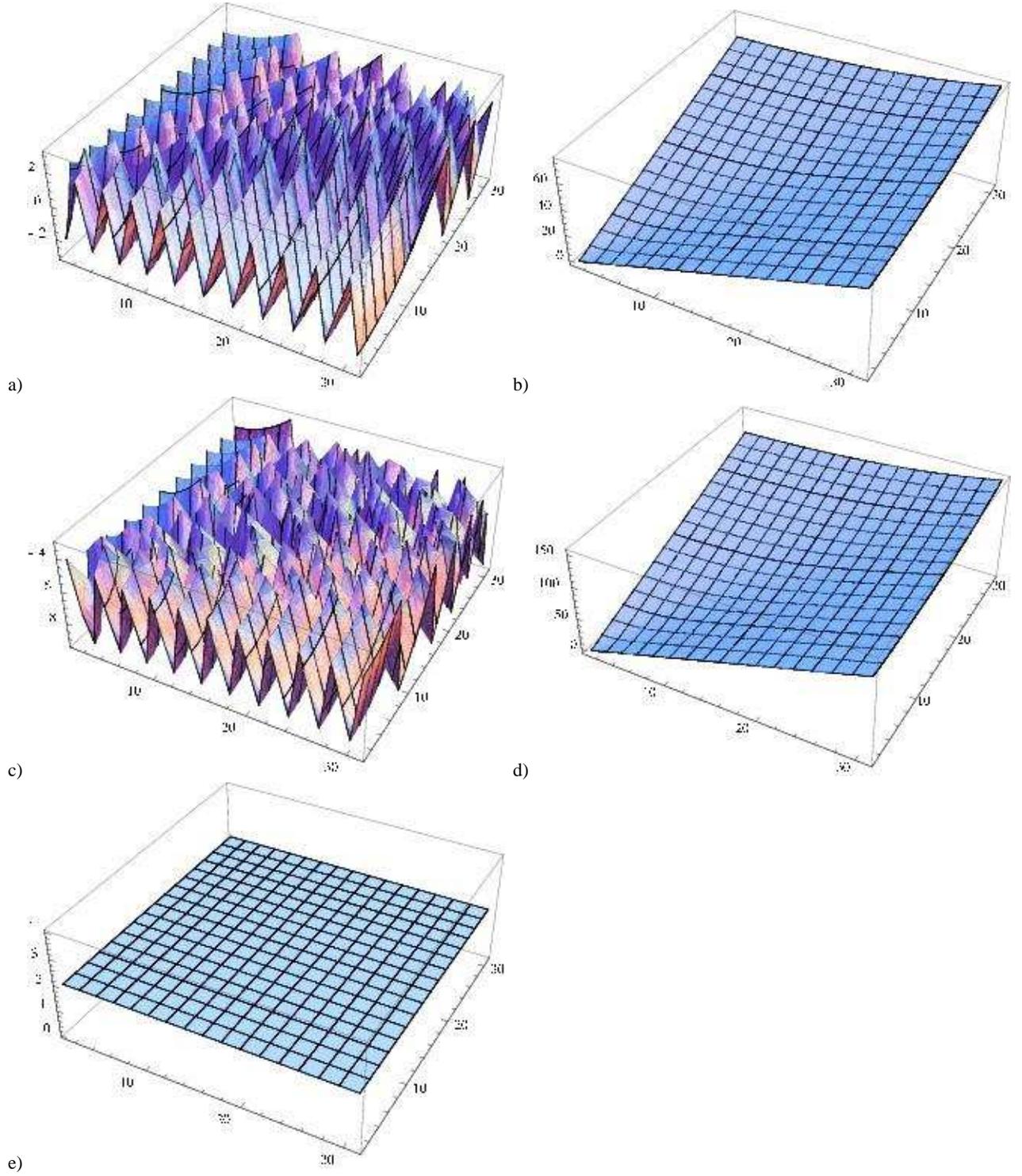}}
    \end{tabular}
    \caption{In these figures, we show the results of the Phase reconstruction algorithm. In a) and c), we show the phase (as
a function of space around one quadrant of the impurity at origin) as evaluated using $Arg(z)$ which restricts the values to one 
branch of the Argument function for $G_0(\RR;\omega)$ and $G_0(\RR;\omega)^2$ respectively. In b) and d), we show the monotonised
phase according to the algorithm described in this section. In e), we confirm that the ratio of the reconstructed phases of 
$G_0(\RR;\omega)^2$ and $G_0(\RR;\omega)$ is identically two everywhere. In this example, $\epsilon(\K) = -2 (Cos[k_x]
+Cos[k_y])$ and $\eta(\K) = 0.25 + 0.1 (Cos[k_x]+Cos[k_y])$ and $\omega = -1$.}
    \label{fig:Phase1}
\end{figure}

Here, we write down the flowchart for the phase reconstruction algorithm that is followed to fix the phase of 
$G_0(\RR;\omega)$ which we get by taking the square root of  the equation 
\begin{equation}
G_0(\RR;\omega)^2 = \frac{G(r,r;\omega) -G_0(\RR=0;\omega)}{G(r=0,r=0;\omega) -G_0(\RR=0;\omega)}*G_0(\RR=0;\omega))^2.
\end{equation}
Upon taking the square root, we get $G_0(\RR;\omega)$ upto a phase of $e^{i \pi}$. To fix this phase, we note that
since the propagator in the continuum $G_0(\RR;\omega)$ has to be a smooth well-behaved function for $\RR \neq 0$ if it is to 
satisfy the Green's function
equations of motion for the Hamiltonian operator and therefore its phase should 
also be a smooth and well-behaved as a function of $\RR$. To see this we start with the equation of motion
for the non-interacting case in the continuum is
\begin{equation}
(i \partial/\partial t + \nabla^2/2m)G_{non}(x,t ; x',t') = \delta(x-x') \delta(t-t')
\end{equation}
which upon Fourier transforming with respect to time gives
\begin{equation}
(\zeta + \nabla^2/2m)G_{non}(x,x';\zeta) = \delta(x-x')
\end{equation}
For $x \neq x'$, the above differential equation has no ill-behaved term and thus $G_{non}(x,x';\zeta)$ has to be 
a well-behaved differentiable function. For the interacting case, the equations of motion is an infinite
hierarchy of differential equations with the successive equations involving higher order Green's functions
(See Vinay Ambegaokar's Chapter on The Green's Function Method in Superconductivity, Vol 1, edited by R. D. Parks).
It is not clear to the author, how one could extend the non-interacting argument to this case. Instead, we argue
as follows. As is usual in perturbation theory, the full propagator in momentum space satisfies a Dyson's equation and is given
by $\tilde{G}(\vec{p},\zeta) = (\zeta - p^2/2m - \Sigma(\vec{p};\zeta))^{-1}$ where $\Sigma(\vec{p};\zeta)$
is called the Self-energy and captures the effect of interactions. If this self-energy doesn't change the analytic
structure of $\tilde{G}(\vec{p},\zeta)$ when compared to $\tilde{G}_0(\vec{p},\zeta)$ (More precisely, the pole
at $\zeta = p^2/2m$ for the non-interaction survives, though it will get shifted off the real axis),
then upon Fourier transforming to real space, the differentiability of $G(x,x';\zeta)$ will be preserved.
Looking at Eq. \ref{eq:contal1}'s continuum version, 
\begin{equation}
\nabla^2 G(\RR;\En) =  \lim_{\delta \rightarrow 0^+} \frac{1}{(2\pi)^2} \int_{B.Z.} 
d \K \frac{|\K|^2 e^{i \K. \RR}}{\En+i\delta- |\K|^2/2m-\Sigma(\K;\En)}
\end{equation}
and if the pole structure of the integrand is same with and without $\Sigma$, then
the differentiabiltity of $G_0(x,x';\zeta)$ implies differentiabiltity of $G(x,x';\zeta)$.
In the case of a lattice, $G_0(\RR;\zeta)$ is well-behaved for $\RR \neq 0$
and at $\RR = 0$ there is a kink in its phase (See the origin in Fig. \ref{fig:Phase1} b)
and d)).

Similarly, $G_0(\RR;\omega)^2$'s phase should also be well-behaved as a function of
space. This is condition that we impose on $G_0(\RR;\omega)^2$ while fixing phases. We start by making a spatial list of 
the phases as given by the $Arg(z)$ function which restricts the phase obtained to the principal branch $(-\pi,\pi]$. Then,
we start at the impurity site for which $\RR = 0$. As we move away from the origin, we multiply phase factors of $e^{i 2 m \pi}$ to
$G_0(\RR;\omega)^2 = |G_0(\RR;\omega)^2| e^{i \phi_{principal}}$ for all $\RR$ and the 
$m$'s are so chosen that if $|\RR'| > |\RR|$ then $\phi'_{principal} + 2 \pi m' > \phi_{principal} + 2 \pi m $.
We implemented
the choosing of $m$'s in the following way :

1) Define a monotoniser function that takes two arguments that lie between $(-\pi,\pi]$ and keeps adding $2 \pi$ to
the second argument till it becomes greater than the first argument. $mono(x,y)$ : Do $y = y + 2\pi$ Till $y > x$.

The following steps are done in each of the symmetry-related octants in space and we write down the steps for
the octant $y > 0$ and $x > y$.

2) Start at origin $(0,0)$. Move a step along x-axis to $(x,y) = (1,0)$. 
Then, $mono(\phi_{principal}(\RR=(x-1,y)),\phi_{principal}(\RR=(x,y))$.

3) Then do $mono(\phi_{principal}(\RR=(x,y)),\phi_{principal}(\RR=(x,y+1))$ along y-direction till $y=x$.

4) Move a step along x-axis. Do $mono(\phi_{principal}(\RR=(x-1,y)),\phi_{principal}(\RR=(x,0))$ where the 
$y$ of the first argument is highest integer such that $x > \sqrt{(x-1)^2 + y^2}$.

5) Repeat step 3) and 4) till the whole octant is covered.

Similar phase fixing is done for all the octants. Once this is done, the phase of $G_0(\RR;\omega)$ is just half that of
the phase-fixed  $G_0(\RR;\omega)^2$. Since, the phase of $G_0(\RR;\omega)^2$ has been made well-behaved,
the phase of $G_0(\RR;\omega)$ will also be well-behaved which is what was desired. In Fig. \ref{fig:Phase1},
we show the result of doing the phase-fixing to numerically calculated $G_0(\RR;\omega)^2$ and also
directly to $G_0(\RR;\omega)$ and find that they are in the correct ratio of two.

\subsection{Proof of self-energy relation}
In this section we prove that the relation between the electron and hole lifetimes, $\eta_{hole}(-\omega)=-\eta_{electron}(\omega)$.
We will do this using the 2x2 Matsubara formalism. In this formalism, the Green's function for the non-lifetime broadened
system in the normal state(i.e. no superconductivity) looks like
\begin{equation}
G_0(k;i \omega_n)^{-1} =
\big( \begin{array}{lr}
			i\omega_n  - \epsilon(k) &  0       \\
			0    & i \omega_n + \epsilon(k)  \\
		\end{array} \big)
\label{eq:freeSCprop1}
\end{equation}
where $\omega_n = (2n + 1)\pi T$ is the fermionic Matsubara frequency.

We will first prove the relation in the case of the normal electrons coupled to phonons. The
self-energy in this case looks like 
\begin{eqnarray}
\label{eq:self-energy-corr}
\Sigma(\vec{k}; i \omega_n) & = & - \frac{T}{N_L} \sum_{\vec{p}, \Omega_m} g(\vec{k}-\vec{q},\vec{q}) g(\vec{k},-\vec{q})
D(\vec{q}; i \Omega_m) \tau_3 G_0 (\vec{k}
-\vec{q}; i \omega_n - i \Omega_m) \tau_3 \nonumber \\
\end{eqnarray}
where $\Omega_m = 2m\pi T$ is the bosonic Matsubara frequency, $\tau_3$ is the third componenet of Pauli matrices in the Nambu space, 
$D(\vec{q}; i \Omega_m)$ is the fourier-transform of the phonon's Green's function $D(\vec{q}; \tau) = - <T_\tau [A(\vec{q};\tau) 
A(-\vec{q};0)]>$ and evaluates to
\begin{equation}
D(\vec{q}; i \Omega_m ) = \frac{1}{2} \big( \frac{1}{i \Omega_m - \Omega_{\vec{q}}}  - \frac{1}{i \Omega_m + \Omega_{\vec{q}}} \big)
\end{equation}
where $\Omega_{\vec{q}}$ is the phonon dispersion. It has the following property : $D(\vec{q}; i \Omega_m) = D(\vec{q}; - i \Omega_m)$.
The $g(\vec{k},\vec{q})$ is the electron-phonon coupling strength coming from the electron-phonon interaction term
\begin{equation}
H_{el-ph} = \frac{1}{N_L} \sum_{\vec{k},\vec{q}, \sigma} g(\vec{k},\vec{q}) c^{\dagger}_{\vec{k} + \vec{q},\sigma}
c_{\vec{k},\sigma} A_{\vec{q}}.
\end{equation}

Using the property $D(\vec{q}; i \Omega_m) = D(\vec{q}; - i \Omega_m)$  and $\Omega_{-m} = -\Omega_{m}$, we can show that(suppressing momenta indices)
\begin{eqnarray}
\Sigma_{22}(- i \omega_n ) & \propto & \sum_{\Omega_m} \frac{D(\Omega_m)}{- i \omega_n - i \Omega_m + \epsilon} \nonumber \\
& = & \ldots + \frac{D(\Omega_{-1})}{- i \omega_n - i \Omega_{-1} + \epsilon} 
+ \frac{D(\Omega_{0})}{ -i \omega_n + \epsilon} + \frac{D(\Omega_{1})}{- i \omega_n - i \Omega_{1} + \epsilon} + \ldots \nonumber \\
& = & \ldots + \frac{-D(\Omega_{-1})}{i \omega_n + i \Omega_{-1} - \epsilon} 
+ \frac{-D(\Omega_{0})}{i \omega_n - \epsilon} + \frac{-D(\Omega_{1})}{i \omega_n + i \Omega_{1} - \epsilon} + \ldots \nonumber \\
& = & \ldots + \frac{-D(\Omega_{1})}{i \omega_n - i \Omega_{1} - \epsilon} 
+ \frac{-D(\Omega_{0})}{i \omega_n - \epsilon} + \frac{-D(\Omega_{-1})}{i \omega_n - i \Omega_{-1} - \epsilon} + \ldots \nonumber \\
& = & - \sum_{\Omega_m} \frac{D(\Omega_m)}{i \omega_n - i \Omega_m - \epsilon} \nonumber \\
& \propto & - \Sigma_{11}(i \omega_n )
\end{eqnarray}
Thus, by analytic continuation, $\Sigma_{22}(- z) = - \Sigma_{11}(z)$ where $\Sigma_{22}$ and $\Sigma_{11}$ are the hole and electron
self-energies respectively. Thus when we analytically continue till $z = \omega + i \delta$ where $\omega$ is real, we see that
$\Sigma_{22}(- \omega - i \delta) = - \Sigma_{11}(\omega + i \delta)$. From the analytic properties of Self-energy
$\Sigma(\vec{p};\omega \pm  i \delta) = \delta\mu(\vec{p};\omega) \mp \frac{i}{2} \eta(\vec{p};\omega)$(see e.g., Eqn. 82 in Vinay 
Ambegaokar's Chapter on The Green's Function Method in Superconductivity, Vol 1, edited by R. D. Parks), we conclude that 
\begin{equation}
\eta_{hole}(-\omega)=-\eta_{electron}(\omega) \ldots QED
\end{equation}
Also, the chemical potential shift is equal for both holes and electrons. This proof can be extended to higher orders in the
electron-phonon coupling by noticing that all higher order terms contributing to self-energy contain odd number of 
fermion propagators, thus allowing the same kind of manipulation done above to go through analogously. This proof extends to
other bosonic modes(e.g. spin wave modes) too since their propagators also satisfy $D(\vec{q}; i \Omega_m) = D(\vec{q}; - i \Omega_m)$. 
This proof also extends to the case of lifetime broadening induced by electron-electron interaction by the same token that 
the self-energy terms always have odd number of fermion propagators.


\begin{thebibliography}{99}
\bibitem{shift}
It is this ``renormalized" dispersion that is measured in FT-STS 
and ARPES.
\bibitem{spectralweight}
Practically, even though the LDOS might not be exactly zero outside
the bandwidth, we saw that if the spectral weight of the spectrum within 
the bandwidth is approximately greater than 0.95, which can be expected
to be true for well-defined quasiparticles, the Kramers-Kroning essentially
works. Moreover, as we show in Fig. \ref{fig:KK}, this issue matters even less
for an energy range much smaller than the bandwidth.
\bibitem{extrapolation}
The knowledge of the dispersion can serve as a guide in this regard. Furthermore, we can have an iterative set-up
where the result of the analysis scheme using the extrapolated full-bandwidth LDOS data can be used 
to make a new and better extrapolation to ensure convergence.
\end{thebibliography}
\end{document}